\documentclass{PoS}

\usepackage[square,comma,numbers]{natbib}
\usepackage{graphicx}
\usepackage{sidecap}

\newcommand{\Jb}{\mbox Jy~beam$^{-1}$}

\newcommand{\cmsixpc}{\mbox{cm$^{-6}$~pc}}
\newcommand{\cmthreepc}{\mbox{cm$^{-3}$~pc}}

\newcommand{\nupeak}{\mbox{$\nu_{\rm peak}$}}

\newcommand{\ttunity}{\mbox{$t_{\tau = 1}$}}
\newcommand{\Msol}{\mbox{M\raisebox{-.6ex}{$\odot$}}}

\title{Recent VLBI Results on SN 1986J and the Possibility of FRBs
  Originating from Inside the Expanding Ejecta of Supernovae}

\ShortTitle{SN 1986J and FRBs}

\author{\speaker{Michael Bietenholz}\thanks{Also at York University, Toronto, Canada.}\\
        SARAO, South Africa\\
        E-mail: \email{michael@hartrao.ac.za}}

\author{Norbert Bartel\\
  York University Toronto, Canada }

\abstract{We discuss our VLA and VLBI observations of supernova 1986J,
  which is characterized by a compact radio-bright component within
  the expanding shell of ejecta.  No other supernova (SN) has such a
  central component at cm wavelengths.  The central component
  therefore provides a unique probe of the propagation of radio
  signals at cm wavelengths through the ejecta of a young SN\@.  Such
  a probe is important in the context of Fast Radio Bursts (FRB),
  which, in many models, are thought to be produced by young magnetars
  or neutron stars.  The FRB signals would therefore have to propagate
  through the expanding SN ejecta.  Our observations of the Type II
  SN~1986J show that its ejecta will remain opaque to cm-wave
  emission like FRBs for at least several decades after the explosion,
  and by the time the ejecta have become transparent, the contribution
  of the ejecta to the dispersion measure is likely small.}

\FullConference{14th European VLBI Network Symposium \& Users Meeting (EVN 2018)\\
		8-11 October 2018\\
		Granada, Spain}

\begin{document}

\section{Introduction}
\label{sintro}

Fast Radio Bursts (FRBs) are bursts of radio emission, 0.1 to 10 Jy at
$\sim$1~GHz, which occur on timescales of milliseconds or less.  They
are characterized by high dispersion measures, DM =
$\int{N_e \cdot dl}$, in the range of 200 to 2000 \cmthreepc.  Their
origin is still mysterious (for recent reviews, see e.g.\
\cite{Katz2016b, Petroff+2016}, and also Marcote {\em this issue}).
These DMs are far higher than expected from our own Galaxy's
interstellar medium.  If the DM is interpreted as being due to the
intergalactic medium, cosmological distances are implied, with
redshifts, $z$ in the range of 0.2 to 1.6\@.  Indeed the single FRB
which has so far been accurately localized, FRB~121102, was found to
be at $z = 0.129$ \cite{Chatterjee+2017}.

Many of the models of FRBs have them originate in young pulsars or
magnetars (see, e.g.\ \cite{Katz2016b}).  A young pulsar or magnetar
would have been born in a supernova (SN) explosion, and therefore
would still be embedded in an expanding cloud of SN ejecta.  The FRB
signals would then have to propagate through the SN ejecta to reach
us.  These ejecta, expected to be partly ionized, are possibly the
source of the large DMs.

\begin{SCfigure}
\centering
\includegraphics[width=0.50\textwidth]{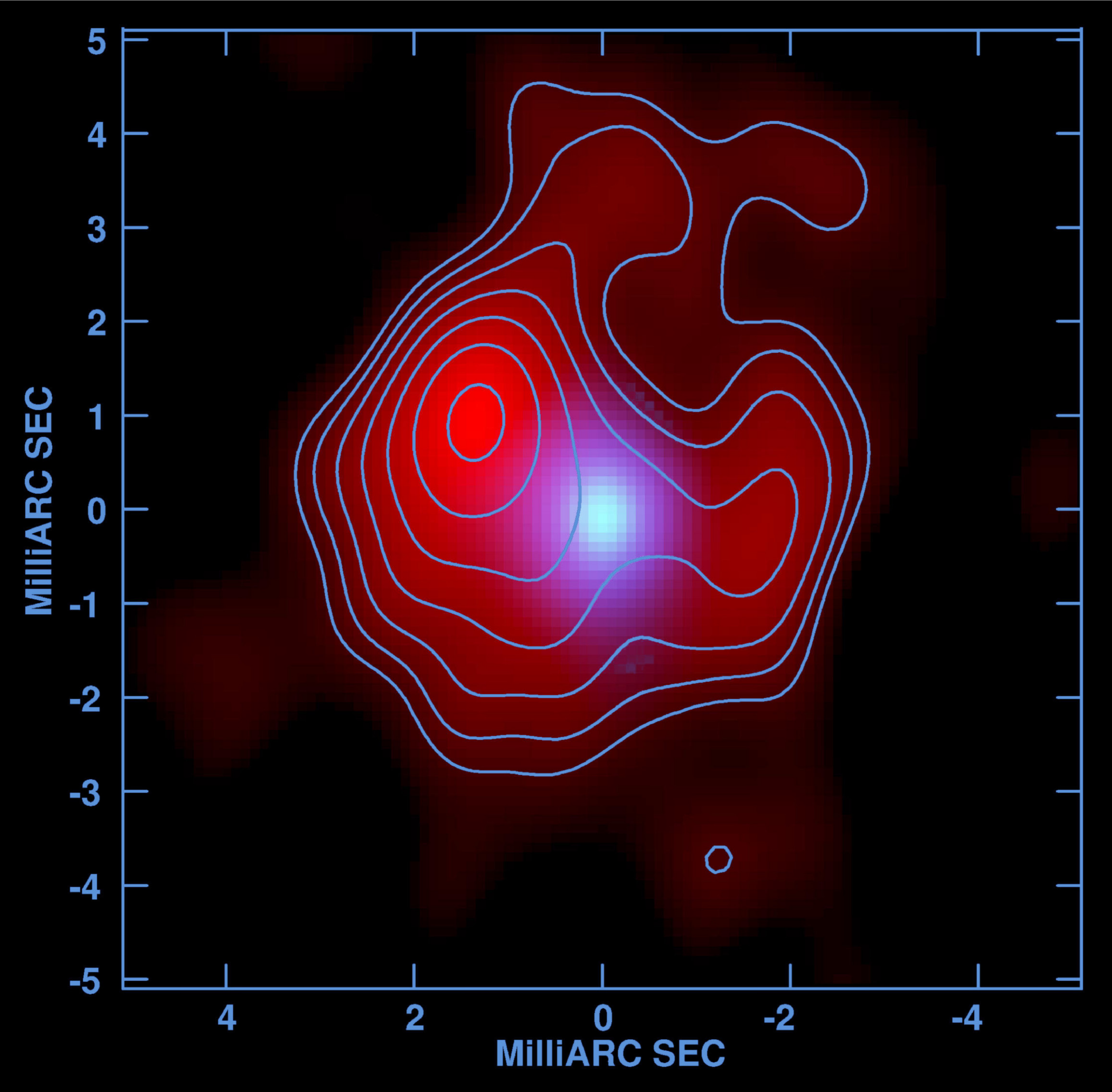}
\caption{A phase-referenced dual-frequency VLBI image of SN~1986J from
  data taken 2002 to 2003, showing the compact, inverted-spectrum
  component located almost precisely in the projected centre of the
  expanding shell.  The red colour and the contours represent the
  5~GHz radio brightness, showing the shell emission.  The contours
  are drawn at 11.3, 16, 22.6, \dots, 90.5\% of the peak 5~GHz
  brightness of 0.55~m\Jb.  The blue through white colours show the
  15~GHz radio brightness distribution, which is dominated by the
  compact, central component which appeared around 1999. North is up
  and east to the left. For details see \cite{SN86J-Sci}.}
\label{fimg}
\end{SCfigure}

It is difficult, however, to obtain any direct observational
constraints on the propagation of radio signals through the ejecta of
a young SN.  SN~1986J is a unique case where we can obtain such
observational constraints.  Its age, at present, is
$t \sim 35$~yr\footnote{Although SN~1986J was first detected in 1986,
  the explosion was a few years earlier, in 1983 \cite{SN86J-1}.}.
It is relatively nearby, in the galaxy NGC~891, at a distance of
10~Mpc (NASA/IPAC Extragalactic Database, NED)\@.  SN~1986J is bright
in the radio, and we have been observing it with very long baseline
interferometry (VLBI) since 1987 and can resolve the expanding ejecta
\cite{SN86J-1, SN86J-2, SN86J-3, SN86J-4, BartelSR1989}.  We show a
dual-frequency VLBI image of SN~1986J in Figure~\ref{fimg}. A singular
feature of SN~1986J, not seen in any other SN at cm wavelengths, is a
compact radio-emitting component which appears to be {\em inside}\/
the expanding shell of ejecta.  This component's central location, its
longevity and its high brightness argue strongly that it is in fact
near the physical centre of the expanding ejecta and not just central
in projection \cite{SN86J-4}.

The nature of this central component is not yet clear.  SN~1986J's
progenitor was sufficiently massive \cite{Rupen+1987} that either a
neutron-star or black-hole compact remnant may have formed.  The
central component may therefore represent radio emission either from a
nascent pulsar-wind nebula or from accretion onto the newly-formed
black hole.  It may, however, also represent shock interaction with a
very aspherical circumstellar medium distribution left over from a
binary progenitor which had undergone a period of common-envelope
evolution. We discuss the various hypotheses in \cite{SN86J-4}.
However, for the present purpose, the physical nature of the central
component is not important, the important thing is only that it
resides {\em within}\/ the expanding cloud of SN ejecta.

No FRB has been observed from SN~1986J, although we note that if one
had occurred, the odds that it would have been detected are quite low.
However, the central component produces radio emission at $\sim$1~GHz,
which would be affected by the intervening ejecta in much the same way
as that from an FRB would be.  What then can we say about the
propagation of FRB-like radio emission through the ejecta of a young
SN on the basis of our observations of SN~1986J?

\section{Observations}
\label{sobs}

We describe our observations briefly here.  More detail can be found
in our series of papers \cite{SN86J-1, SN86J-2, SN86J-3, SN86J-4}.  We
have observed SN~1986J with the Jansky Very Large Array (VLA), to
measure the evolving total flux density at a wide range of
frequencies.  These observations allow us to determine the evolution
of the radio spectral energy distribution (SED) with time.  Other VLA
observations were also obtained earlier by \cite{WeilerPS1990}.  We
further obtained VLBI observations at a number of epochs between 1987
and 2014, mostly using a global array consisting of antennas from the
European VLBI Network (EVN) and the National Radio Astronomy
Observatory (NRAO; USA).  The most recent VLBI observations were
obtained in 2014 Oct., at 5.0 GHz using a global array of EVN and NRAO
antennas (see \cite{SN86J-3}).

\section{Evolution of the Spectral Energy Distribution}
\label{sSED}

\begin{SCfigure}[][th]
\centering
\includegraphics[width=0.6\textwidth]{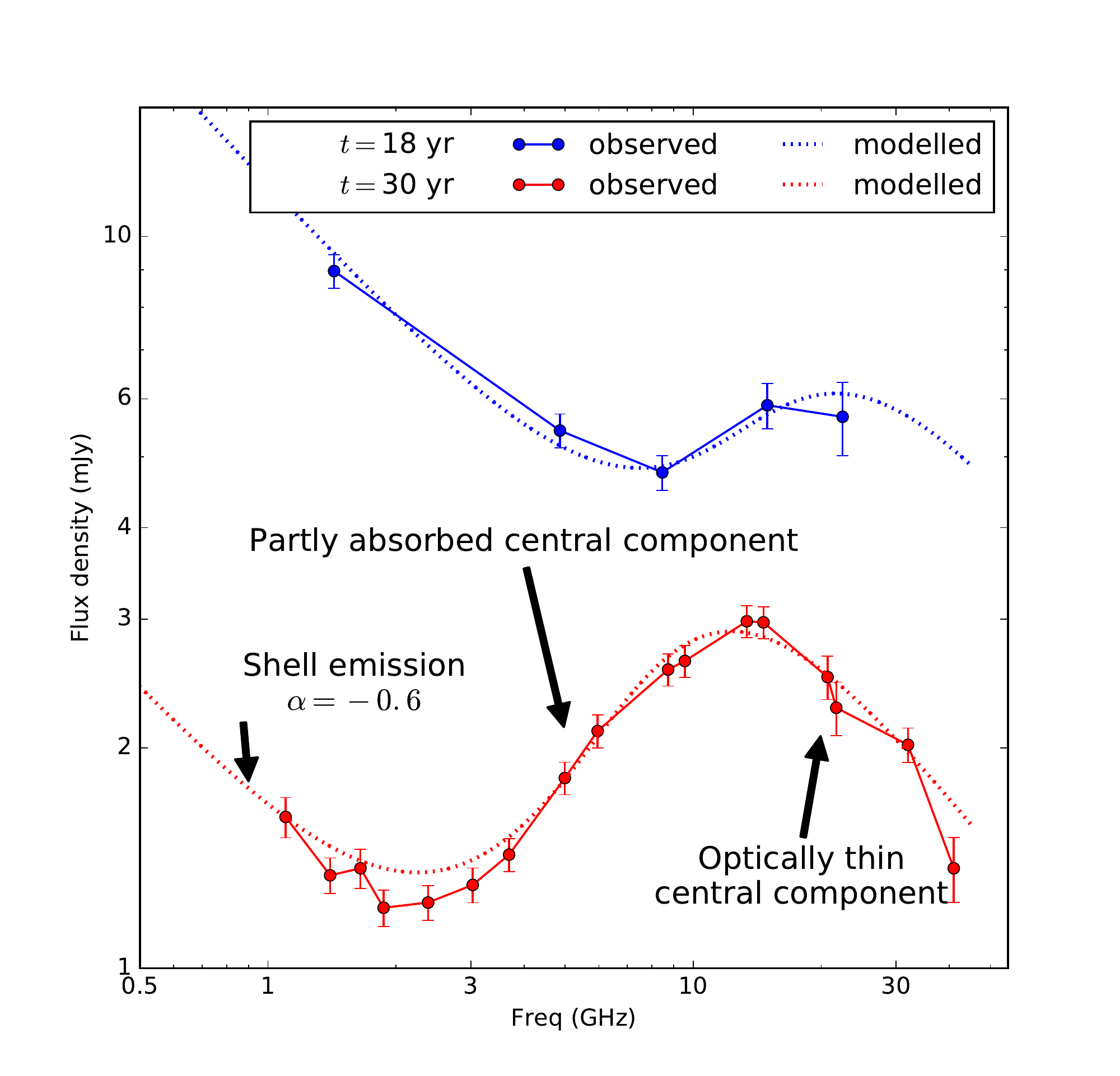}
\caption{The evolving SED of SN 1986J\@. We show only two example
  SEDs, one from $t = 18$~yr and the latest one from $t=30$~yr.  For
  SEDs at intermediate times showing the evolution in more detail, see
  \cite{SN86J-4}.  We indicate the parts of the SED due to the central
  component and to the optically-thin shell on the lower
  curve.  The uncertainties are estimated standard errors, with
  statistical and systematic contributions added in quadrature.  The
  dotted lines show our Bayesian fit of a simple two-component model
  (one for the shell, one for the central component) to the evolving
  SED\@. See text and \cite{SN86J-4} for details.}
\label{fSED}
\end{SCfigure}

We show the evolution of the SED in Figure~\ref{fSED}.  At low
frequencies, the SED has the form of the simple powerlaw typically
seen for SNe, with spectral index, $\alpha \simeq -0.5$ (where
$S_\nu \propto \nu^\alpha$, where $\nu$ is the frequency).  This is
the radio emission due to the expanding shell, which is optically
thin.  At high frequencies, the radio emission is dominated by that of
the central component, and is still partly absorbed.  An inversion
appears in the SED at the frequency where the central component is
becoming prominent.  At yet higher frequency, the central component
also becomes optically thin, and the spectrum again turns over.  The
two inflection points (inversion point and high-frequency turnover)
both move downward in frequency as a function of time, as can be seen
in Fig.~\ref{fSED}.

In the latest (bottom) SED curve in Fig.~\ref{fSED}, at $t = 30$~yr,
the emission from the central component is absorbed below a turnover
frequency of $\nupeak \simeq 15$~GHz, with the absorption being due to
the free-free mechanism \cite{SN86J-1}.  At $t = 30$~yr, therefore,
any signal at 1~GHz, such as an FRB, would still be strongly absorbed
by the SN ejecta.  If FRBs originate in young SNe similar to SN~1986J,
we would expect not to see them for at least several decades after the
explosion.

We can estimate the time until transparency at 1 GHz for SN~1986J as
follows.  The optical depth to free-free emission is given by:
$$\tau    =  3.28 \times 10^{-7} \, 
  \left(\frac{\nu}{\mathrm{GHz}}\right)^{-2.1} \, 
  \left(\frac{T_e}{10^4 \, \mathrm{K}}\right)^{-1.35}\, 
  \left(\frac{\mathrm{EM}}{\cmsixpc}\right),$$
where $T_e$ is the electron temperature, which we take to the
$10^4$~K, and EM $ = \int{N_e}^2 dl,$ is the emission measure, where
$N_e$ is the number density of free electrons, and $l$ is the
path-length along the line of sight from the centre of the SN to the
observer.

The turnover frequency is approximately that where $\tau_\nu =
1$\@. For SN~1986J at $t = 30$~yr, \nupeak = 15~GHz.  We can then
calculate that EM $\simeq 9\times10^8$~\cmsixpc.  The SED however is
not static, but rather is evolving in time. In particular \nupeak\ is
decreasing with time as can be seen in Fig.~\ref{fSED}.  To
characterize the evolving SED of SN~1986J, we fitted a model with 8
free parameters to the flux density measurements (such as those in
Fig.~\ref{fSED}) between $t = 14$ and 30 yr using Bayesian statistics.

Our model has two parts: $(a)$ the shell, which is optically thin and
whose flux density is a power-law function of time, and $(b)$, the
central component, whose intrinsic flux density is also a power-law
function of time, but which suffers time-dependent absorption by
material whose emission measure, EM, is a power-law function of
time.  Both parts are assumed to have an intrinsic spectrum (before
any absorption) of power-law form.  The eight free parameters of the
model are: the two intrinsic flux densities and the EM at $t = 20$~yr,
the two optically-thin spectral indices, and the powerlaw index of the
time dependencies of the two flux densities and of the EM with time.

For details of the fitted values, and other results, please see
\cite{SN86J-4, SN86J-FRB}.  Of interest in our context of FRB signals
propagating through the ejecta is the solution for EM:

$$ {\rm EM} = (1.64 \pm 0.21) \times 10^9 \, (t/{\rm \, 20 \,
  yr})^{-2.72 \pm 0.26} \; \cmsixpc .$$
The EM, and thus the absorption, are decreasing with time.  This is
expected as the SN is expanding. In a homologously expanding system
with a constant number of free electrons and radius, $r \propto t^m$,
EM would be $\propto t^{-5m}$. We determined the expansion rate of
SN~1986J from the sequence of VLBI images and found that $m =
0.69\pm0.03$ \cite{SN86J-2}.  The time-dependence of the EM implies a
slightly more decelerated expansion with $r \propto t^{-0.54 \pm
  0.05}$, suggesting perhaps some on-going fragmentation in the
ejecta, which would cause the opacity to not scale simply with radius.

Extrapolating our fitted time-dependence of EM, we can calculate the
age at which the 1-GHz optical depth would be unity, $\, \ttunity =
200$~yr.  Given that our model of the SED was only approximate, the
extrapolation from the observations at $t=30$~yr to 200~yr is rather
uncertain.  However, given the present high optical depth at 1~GHz, it
is clear that the ejecta would not become transparent at 1~GHz for at
least several decades from now, so we can probably set a lower bound
of $\ttunity > 60$~yr.

At $t=20$~yr, the fitted EM is $1.64 \times 10^9 $~\cmsixpc.  At that
time the radius of SN~1986J, $r_{\rm SN} = 6 \times 10^{17}$~cm, and
the radius of the reverse shock was $\sim 4.3 \times 10^{17}$~cm
\cite{SN86J-2}. If we take the free electrons to be uniformly
distributed, we can calculate a total ionized mass, $M_{\rm ion}
\gtrsim 40$~\Msol.  Such a high value of $M_{\rm ion}$ is not
reasonable.  We can conclude therefore that the ionized absorbing
material must be non-uniformly distributed.

A non-uniform distribution of $M_{\rm ion}$ is not unexpected: in a
SN, a dual shock structure will form with a forward shock being driven
into the CSM and a reverse shock being driven back into the ejecta,
with the thickness of the region between the shocks being $\sim$20\%
of the forward-shock radius (e.g.\ \cite{Chevalier1982b}).  The
material between the two shocks is at sufficiently low density and
high temperatures ($T_e > 10^6$~K; \cite{Chevalier1982b,
  LundqvistF1988}) that it does not contribute significantly to the
radio absorption, or to the EM or DM \cite{SN86J-FRB}.

The unshocked ejecta interior to the reverse shock are expected to be
ionized by the SN shock breakout, but may recombine fairly rapidly
thereafter.  The radiation from the two shocks is likely to keep some
region towards the outer radius ionized.  In the specific case of
SN~1986J, radiation from the central component, whatever its physical
nature, may also keep some region near the centre ionized.  With
reasonable values of $M_{\rm ion} \leq 5$~\Msol, EM = $1.64 \times
10^9$~\cmsixpc\ can be achieved either with a region near the centre
with a radius $< 0.4 \, r_{\rm SN}$, or with a thin shell of ionized
material near the reverse shock, with thickness $0.002\, r_{\rm SN}$.
We consider the concentration of $M_{\rm ion}$ in such a thin shell
unlikely, so it is more likely that most of the absorption comes from
dense ionized ejecta near the centre of the SN.

One of the distinguishing features of FRBs are their high DMs.  Can we
say anything about the DM produced as signals propagate through SN
ejecta?  At our extrapolated time of transparency at 1~GHz, $\ttunity
\simeq 200$~yr, the EM would be $3 \times 10^6$~\cmsixpc, and the
extrapolated $r_{\rm SN}$ would be $\sim$0.9~pc \cite{SN86J-2}.
For a shell of material ionized by the shocks, the shell must be very
thin to be of reasonable mass ($<5$~\Msol), and the DM remains
$<50$~\cmthreepc.  Larger DMs would occur if the ionized material is
near the centre of the SN, ionized by some central source.  In this
case, an ionized region of $M_{\rm ion} = 5$~\Msol\ of radius
$\sim 0.2\;r_{\rm SN}$ or $\sim 0.2$~pc, would produce a DM
$\sim$800~\cmthreepc, which is in the range observed for FRBs (more
details on the calculation of the DM can be found in \cite{SN86J-FRB}.

\section{Conclusions}

On the basis of our VLBI and VLA observations of the central component
of the Type II SN~1986J, we can conclude that it is still opaque to 1
GHz radiation 30 yr after the explosion, and will likely remain so for
at least another 30 yr.  Therefore, If FRBs do originate in young
neutron stars or magnetars from SNe like SN~1986J, we would not see
them until at least several decades after the SN explosion.  By the
time the ejecta {\em have}\/ become transparent at 1~GHz, the
contribution of the ejecta to the DM is likely only a fraction of the
values typically seen for FRBs, unless the ejecta are ionized from the
centre.

Looking ahead, the Square Kilometre Array will massively boost the
sensitivity of VLBI observations.  It will therefore allow VLBI
observations of SNe to be extended much longer than is possible at
present, and will likely make it possible to resolve SNe at ages of
several decades when they have become transparent to FRBs.
The possibility of imaging FRB-engines is an exciting
prospect for VLBI with the SKA.

\section*{Acknowledgements}

The author acknowledges financial support from the ``VLBI with the
SKA'' work package is part of the Jumping JIVE project, that has
received funding from the European Union's Horizon 2020 research and
innovation programme under grant agreement No 730884.

\newcommand{\amp}{\&}
\newcommand{\araa}{Ann. Rev. Astron. Astrophys.}
\newcommand{\aap}{Astron. Astrophys.}
\newcommand{\aapr}{Astron. Astrophys. Rev.}
\newcommand{\aaps}{Astron. Astrophys. Suppl. Ser.}
\newcommand{\aj}{AJ}
\newcommand{\apj}{ApJ}
\newcommand{\apjl}{ApJL}
\newcommand{\apjs}{ApJS}
\newcommand{\apss}{ApSS}
\newcommand{\baas}{BAAS}
\newcommand{\memras}{Mem. R. Astron. Soc.}
\newcommand{\memsai}{Mem. Soc. Astron. Ital.}
\newcommand{\mnras}{MNRAS}
\newcommand{\iaucirc}{IAU Circ.}
\newcommand{\jrasc}{J. R. Astron. Soc. Can.}
\newcommand{\nat}{Nat}
\newcommand{\pasa}{PASA}
\newcommand{\pasj}{PASJ}
\newcommand{\pasp}{PASP}
\newcommand{\prd}{Phys. Rev. D}
\newcommand{\prl}{Phys. Rev. Lett.}
\newcommand{\ssr}{Space Sci. Rev.}

\bibliographystyle{JHEPv2.7}
\footnotesize

\bibliography{mybib1}

\end{document}